\documentclass[final,5p,times,twocolumn]{elsarticle}

\usepackage{lineno,hyperref}
\usepackage{setspace}
\modulolinenumbers[5]

\journal{Nucl. Instr. Meth. Phys. Res. A}

\begin{document}

\begin{frontmatter}

\title{A Database for Storing the Results of Material Radiopurity Measurements}

\author[sjtu,lbnl]{J.C.~Loach}
\author[smu]{J.~Cooley}
\author[kit]{G.A.~Cox}
\author[sjtu]{Z.~Li}
\author[lbnl]{K.D.~Nguyen}
\author[lbnl]{A.W.P.~Poon}

\address[sjtu]{Department of Physics and Astronomy, Shanghai Jiao Tong University, Shanghai 200240, China}
\address[lbnl]{Nuclear Science Division, Lawrence Berkeley National Laboratory, Berkeley, CA 94720, USA}
\address[smu]{Department of Physics, Southern Methodist University, Dallas, TX 75275, USA}
\address[kit]{Institute for Nuclear Physics, Karlsruhe Institute of Technology, 76131 Karlsruhe, Germany}





\begin{abstract}

Searches for rare nuclear processes, such as neutrinoless double beta-decay and the interactions of WIMP dark matter, are motivating experiments with ever-decreasing levels of radioactive backgrounds. These background reductions are achieved using various techniques, but amongst the most important is minimizing radioactive contamination in the materials from which the experiment is constructed. To this end there have been decades of advances in material sourcing, manufacture and certification, during which researchers have accumulated many thousands of measurements of material radiopurity. Some of these assays are described in publications, others are in databases, but many are still communicated informally. Until this work, there has been no standard format for encoding assay results and no effective, central location for storing them. The aim of this work is to address these long-standing problems by creating a concise and flexible material assay data format and powerful software application to manipulate it. A public installation of this software, available at \url{http://www.radiopurity.org}, is the largest database of assay results ever compiled and is intended as a long-term repository for the community's data.

\end{abstract}


\end{frontmatter}


\section{Introduction}

Searches for rare nuclear processes have led to many of the most important results in nuclear and particle physics over recent decades \cite{sno, kamland, gerda, cdms} and offer the prospect of many more in years to come \cite{lux, mjd2, cuore}. Crucial to these experiments, and especially to those that search for rare weak interactions, such as neutrinoless double beta-decay and the interactions of WIMP dark matter, is the suppression of events due to radioactive backgrounds. This can be achieved through experimental design or analysis, but the most basic strategy is to directly address the source terms by shielding the experiment from environmental radioactivity and, of particular relevance here, constructing the experiment from low-radioactivity materials.

Selecting candidate low-radioactivity materials is as much an art as it is a science, relying on inference from previous measurements and on the instinct of the experimentalist. It is also a critical task because the testing and certification of candidate materials, though a well-established process, involves precision measurements that consume significant amounts of time, cost and effort. Candidates must be selected judiciously and how well this can be done depends upon the quality of information available to the researcher. Of particular importance is their access to previous measurements of similar materials.

A central repository of material radiopurity measurements is strongly motivated by these considerations, and the authors are not the first to propose one or, even, to build one. Previous efforts such as the public material assay database created by the ILIAS collaboration \cite{ILIAS} have been invaluable to the low-background physics community. But despite their usefulness, none have fully met the community's needs. There have been issues with the limited scope of the data sets and difficulties in augmenting them, as well as with non-portability of data and difficulties in querying it. Solving these issues was the motivation for this work, and we describe ways of storing and handling material assays that are intended to address them.

Our work comprises two parts: a data format in which assays can be encoded and a piece of software for manipulating the encoded data. In a practical sense these components are intertwined, but there is a formal separation between them that is reflected in the organization of this paper. Our material assay data format is presented in the next section, free of association with any particular database software. Following this we discuss our web application Persephone \cite{persephone}, which is a tool for storing, viewing and manipulating data encoded in the format. And, finally, we describe a specific installation of our software that we intend as the central repository for the community's data.

\begin{figure}
  \begin{center}
  \includegraphics[scale=0.8]{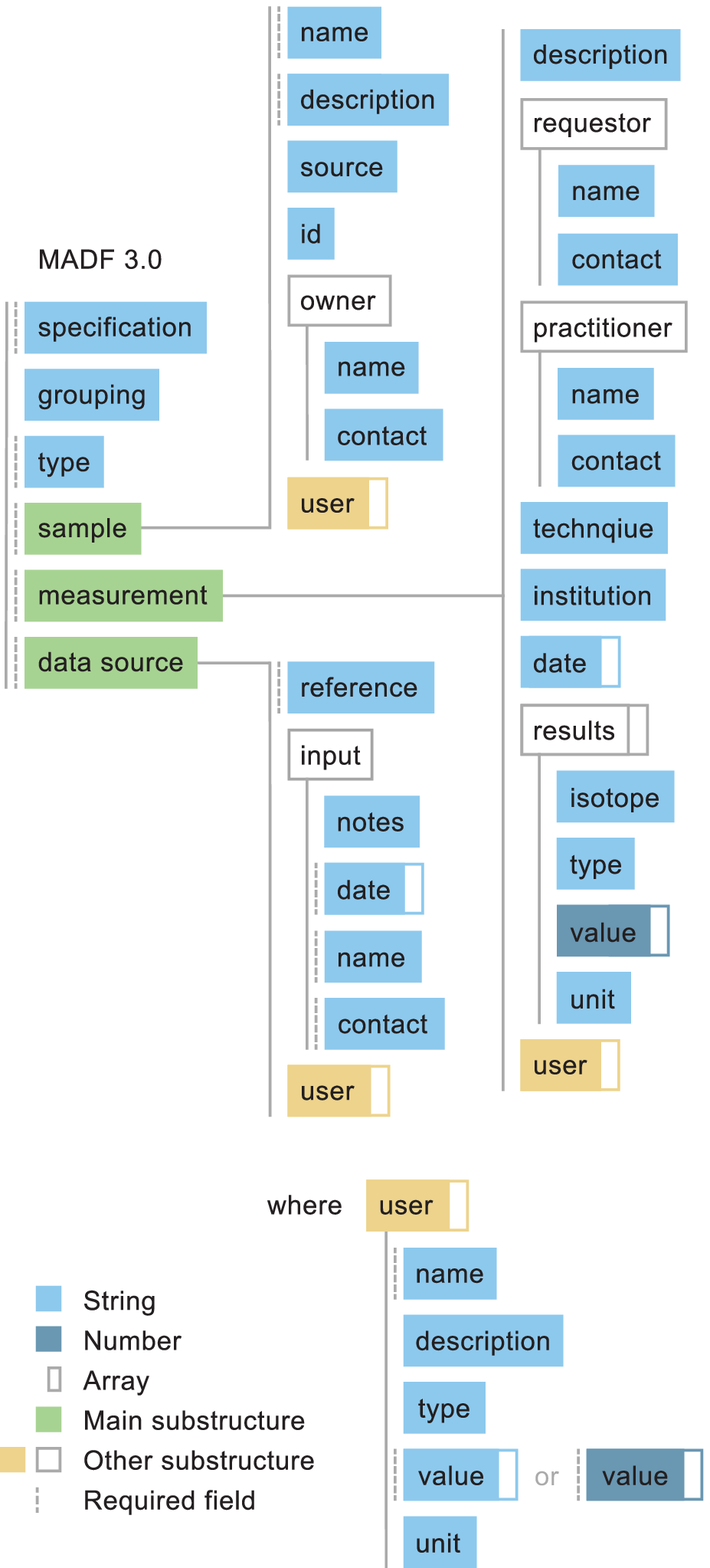}
  \end{center}
  \caption{The structure of the MADF 3.0 data format. Each entry in a \texttt{user} array has the structure given in the lower-right part of the figure. These arrays are used to add extra, use-case-specific fields to the core data structure. The data format is expressed in JSON. Further details are given in Table \ref{t:schema}.}
  \label{f:schema}
\end{figure}

\begin{table}
\caption{The structure of the MADF 3.0 data format. All fields are strings unless stated otherwise. Fields marked with a $\ast$ are required.}
\begin{singlespace}
\small
\begin{center}
\begin{tabular}{ l   p{5cm}  }
\hline
Field & Description \\
\hline
\texttt{specification$\ast$} & MADF specification version. \\
\texttt{grouping} & Experiment name or similar. \\
\texttt{type$\ast$} & Fixed value \texttt{"assay"} indicating the document type.\\

\hline

\multicolumn{2}{l}{\texttt{sample}} \\
\enskip\texttt{.name$\ast$} & Concise description.\\
\enskip\texttt{.description$\ast$} & Detailed description. \\
\enskip\texttt{.id} & Identification number. \\
\enskip\texttt{.source} & Where the sample came from. \\
\enskip\texttt{.owner} & Who owns the sample. \\
\enskip\enskip\texttt{.name} & Name.  \\
\enskip\enskip\texttt{.contact} & Email address or telephone no.\\
\enskip\texttt{.user} & User field array. See below.\\

\hline

\multicolumn{2}{l}{\texttt{measurement}} \\
\enskip\texttt{.description} & Detailed description.\\
\enskip\texttt{.requestor} & Who coordinated the measurement. \\
\enskip\enskip\texttt{.name} & Name. \\
\enskip\enskip\texttt{.contact} & Email address or telephone no. \\
\enskip\texttt{.practitioner} & Who did the measurement. \\
\enskip\enskip\texttt{.name} & Name. \\
\enskip\enskip\texttt{.contact} & Email address or telephone no. \\
\enskip\texttt{.technique} & Technique name.\\
\enskip\texttt{.institution} & Institution name.\\
\enskip\texttt{.date} & Array of strings. See Appendix C. \\

\enskip\texttt{.results} & Array of measurements.\\
\enskip\enskip\texttt{.isotope} & Isotope name, usually in the format \texttt{symbol-mass number}.\\
\enskip\enskip\texttt{.type} & \texttt{measurement}, \texttt{limit} or \texttt{range}. \\
\enskip\enskip\texttt{.value} & Array of one to three numbers.  See Appendix A. \\
\enskip\enskip\texttt{.unit} & Unit chosen from a restricted set of choices. See Appendix B. \\
\enskip\texttt{.user} & User field array. See below. \\

\hline

\multicolumn{2}{l}{\texttt{data\_source}} \\
\enskip\texttt{.reference$\ast$} & Where the data came from.\\
\enskip\texttt{.input} & Data entry details.\\
\enskip\enskip\texttt{.notes} & Input simplifications, assumptions. \\
\enskip\enskip\texttt{.date$\ast$} & Array of strings. See Appendix C. \\
\enskip\enskip\texttt{.name$\ast$} & Name. \\
\enskip\enskip\texttt{.contact$\ast$} & Email address or telephone no. \\
\enskip\texttt{.user} & User field array. See below. \\

\hline

\multicolumn{2}{l}{\enskip\texttt{.user}} \\
\enskip\enskip\texttt{.name$\ast$} & Concise description, usually a single word. \\
\enskip\enskip\texttt{.description} & Detailed description.\\
\enskip\enskip\texttt{.type} & \texttt{measurement}, \texttt{limit}, \texttt{range} or \texttt{string}.\\
\enskip\enskip\texttt{.value$\ast$} & String or an array of one to three numbers. See Appendix A.\\
\enskip\enskip\texttt{.unit} & Unit. See Appendix B.\\

\hline

\end{tabular}
\end{center}
\label{t:schema}
\end{singlespace}
\end{table}

\section{The Material Assay Data Format}

The Material Assay Data Format (MADF) describes a way to encode assays of material radiopurity in JavaScript Object Notation (JSON) \cite{json}. JSON is a language-independent, open standard for encoding structured data. It is especially widespread in Internet communications and is natively read by many advanced and popular computational tools, including those used in physics analyses.

Each MADF JSON document represents a single assay performed on a particular sample of material. It contains three main substructures: \texttt{sample} describes the sample of material being assayed; \texttt{measurement} describes the assay process and its results, including a list of isotopes with their measured values or limits; and \texttt{data\_source} gives the origin of the information encoded in the document and describes how it was encoded. Overviews of MADF 3.0 are shown in Figure \ref{f:schema} and Table \ref{t:schema}. Further details are given in the appendices.

The design principles behind the data format are summarized as follows:

\begin{description}

\item[Simplicity.] Each assay is encoded in a single, semi-structured, human-readable JSON document and there are no structural relationships between documents. In contrast to relational databases, with their arrangements of tables, all information is encapsulated in a single piece of data. This makes the data highly portable, and assays can even be output and stored in self-contained text files. 

In MADF many fields can store extended, descriptive information and only a small number of fields are mandatory. In this way it is more similar to a structured experimental report than, for example, a line in a spreadsheet. This simplicity helps the process of encoding data because there is no need to parse data from disparate sources into a highly-structured form. It helps makes the process amenable to automation and minimizes the decisions required during data entry (and therefore the required skill level of the people doing the entry). One consequence of a simple data format is that it tends to make querying data more difficult but this problem is eased by modern tools such as MapReduce \cite{mapreduce} which are implemented in databases designed to store JSON documents.

\item[Inclusivity.] MADF tries to minimize constraints on what can be entered in particular fields. For example, there are no requirements that measured values have uncertainties or that limits have confidence levels. The rationale is that more information is generally better, and that the intended users will be sufficiently qualified to judge what is relevant for their application. A measured value without an explicit uncertainty might not be appropriate in a scientific publication but it is not without meaning in certain contexts. However, expecting the user to make judgements only works when the level of information provided is sufficient and, in particular, when they can understand the context and origin of the data. MADF therefore contains required fields for storing a reference to the source data and information on data entry, including the contact details of the person who entered the data. There is also an optional field for describing any assumptions or simplifications made during data entry.

\item[Flexibility.] The core of MADF is a simple, compact structure, but it can be extended arbitrarily to accommodate specific needs. A typical use case might involve extra fields for information that is important to people who perform assays, but less so to those who use the results. Examples might be file names, calibration data, images, energy spectra or even analysis code (stored as serialized objects or, in the case of Python, as IPython Notebook JSON objects). The data structure can be extended in two ways: freely or through \texttt{user} arrays. Free extension means adding arbitrary fields anywhere within the structure. In contrast, \texttt{user} arrays contain predefined structures that can be added only to the three main \texttt{sample}, \texttt{measurement} and \texttt{data\_source} substructures. The difference between them is that all applications implementing MADF should recognize and display \texttt{user} arrays, while there are no requirements on how they should handle free extensions. Our application Persephone, discussed below, ignores free extensions.

\end{description}

The structure of MADF is described by a JSON schema \cite{jsonschema}, against which documents can be validated using open source libraries (such as \cite{jsonvalidate} for Python). Future development of MADF is encouraged by the independence of the schema from the data representation, which means that changes can be made with substantial freedom and with few consequences for existing data. One reason for the failure of previous databases has been their structural rigidity, which has made it difficult to add new data and difficult to adapt to new forms of data. A database using JSON documents is maximally flexible and adaptable to future needs.

JSON documents can be stored in a variety of database systems. These include databases specially designed around the format, as well as traditional relational databases. We discuss one database system in the following section in the context of our web application.

\begin{figure}
  \begin{center}
  \includegraphics[scale=0.8]{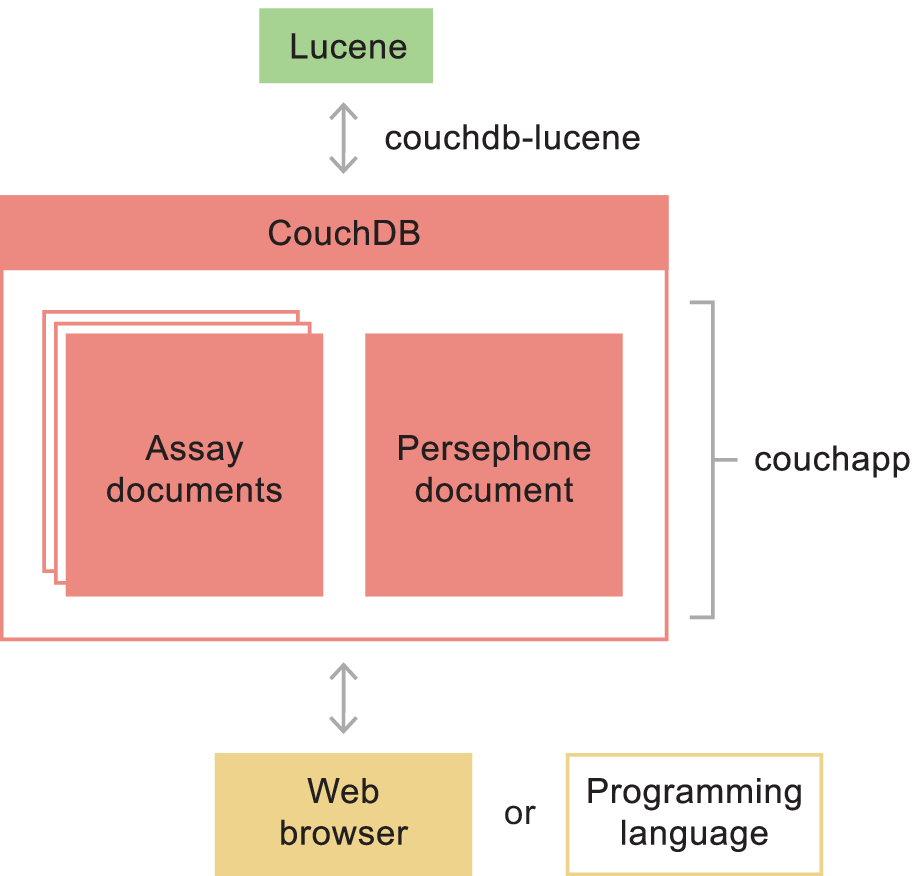}
  \end{center}
  \caption{The structure of the database and web application. Gray arrows indicate HTTP requests.}
  \label{f:software}
\end{figure}

\section{Persephone}

Persephone \cite{persephone} is an open-source software package for storing, displaying and manipulating assays encoded in MADF. It is a JavaScript-based web application stored within, and served by, an Apache CouchDB database \cite{couchdb}. The architecture is illustrated in Figure \ref{f:software}. 

CouchDB is an open-source, document-oriented database which uses JSON documents for data storage. It is schema-less, meaning that it does not enforce any pre-defined structure on its documents or on the relationships between them; schemas are specified and enforced by user-developed code that interacts with the database. CouchDB uses MapReduce as its query language and HTTP as its primary API. It runs as a web server and provides sophisticated functionality for replicating databases between different locations and guaranteeing eventual consistency. CouchDB does not currently have native search capabilities but it can interface with other software such as Apache Lucene \cite{lucene, couchdblucene} to provide this functionality. 

These features make CouchDB an ideal tool for a material assay database (and, indeed, for a variety of applications in physics experiments \cite{cox, mjparts}). It provides the flexibility required by the data format and its use of HTTP makes it easy to interface with using most programming languages. In addition, being both schema-less and a web server, CouchDB has the ability to store entire client-side web applications as documents within the database and serve them to users. Tools such as couchapp \cite{couchapp} are available to help encode web applications as JSON documents. Persephone is an application of this type.

Persephone is a client-side web application stored inside a CouchDB database and served by it. It is built from HTML, CSS and JavaScript using standard web programming techniques, such as AJAX, and standard libraries, such as jQuery \cite{jquery}, jQueryUI \cite{jQueryUI} and the CouchDB jquery plugin. The user interface is a single web page containing tabs for search, data entry, data editing, settings and user login. 

The search interface allows standard search grammar such as wild cards, logical operators and named fields. Search results are initially shown in an abbreviated form, where they can be sorted or removed from the display. The user can expand the abbreviated forms to two deeper levels of detail. Each entry has a tools menu that allows the user to report errors, export the entry (in JSON, CSV, XML or HTML) or, if they have appropriate permissions, invoke the data editing interface.

The data entry interface is a HTML form that allows assays to be entered in a user-friendly way. It supports the whole MADF specification with the exception of free extensions (the addition of fields outside the core data structure and user arrays). Once submitted, data is validated against the JSON schema and then stored in the database. The interface for data editing is similar in appearance and function. 

The settings tab allows the user to set persistent values for the email address used to report errors and for the number of search results that are displayed by default. 

The login tab provides an interface to CouchDB's user authentication features, which allow simple management of usernames, passwords and read/write permissions.

Persephone is accompanied by a set of python tools that provide command line functionality for submitting, downloading, editing and validating documents. Python provides excellent support for JSON using the json module. The jsonschema module is used to handle validation and the couchdb module is used to communicate with the database. These tools allow data to be downloaded into text files and uploaded from them, and provide an alternative way to edit existing data and submit new data. The scripts can be incorporated into larger codes to assist automatic conversion of data from other sources.

The application is designed to be run in a number of different contexts: by private individuals, by institutions who perform material assays, by experimental collaborations, and also as a central public database of community data. It can be installed on local instances of CouchDB, on public-facing servers, or in the cloud using commercial services, with databases easily replicated between installations. It could be adapted to run on mobile devices. This model allows an experiment or institution to easily share their internal data with the wider community (by replicating the contents of their local database to the public database). The distributed nature of this model should also help protect the community's data by preventing reliance on one individual or institution as a custodian.

\section{The Community Database}

A public instance of Persephone exists at \url{http://www.radiopurity.org}. This database is hosted and managed by SNOLAB and is intended to be a long-term repository for the community's assay data. It currently holds more than 1000 assays, including measurements from peer-reviewed publications \cite{Leonard2008490, Aprile2012573, Alimonti1998141, mjassay} and the ILIAS database. Published data are added as the authors become aware of them (there is a contact form on the website) and ways to incorporate more historical data are being explored. Encoding historical data can be a substantial undertaking and the amount of this data that ends up in the database will be strongly dependent on the availability of resources. We encourage the community to support this valuable activity.

\section{Conclusion}

MADF, Persephone and the community database are three components of a long-term solution to the problem of effectively storing, sharing and searching the results of material assays. MADF provides a flexible, human-readable, non-propriatorey data format for encoding material assays as JSON documents. These documents can be stored in a CouchDB database or easily converted to other formats for use elsewhere. Persephone is an open-source web application for manipulating assays encoded in MADF. It is stored inside a CouchDB database, which also serves it to the user. Persephone allows data to be searched, submitted and edited in a powerful and convenient way. The community installation of Persephone at \url{http://www.radiopurity.org} is intended as the long-term repository for the community's assay data.

\section{Acknowledgements}

This work was supported by: the U.S. Department of Energy, Office of Science, Office of Nuclear Physics, under Contract No. DE-AC02-05CH11231; by the Shanghai Key Lab for Particle Physics and Cosmology (SKLPPC), Grant No. 11DZ2260700; by the Helmholtz Alliance for Astroparticle Physics (HAP) funded by the Initiative and Networking Fund of the Helmholtz Association; and by the National Science Foundation under Grant No. PHY 124260 awarded to the AARM (Assay and Acquisition of Radiopure Materials) collaboration. We acknowledge the support of IBM Cloudant in hosting an early version of our community database and to SNOLAB for their ongoing hosting and management of the site. We acknowledge support and guidance from colleagues in the underground science community, including from the AARM, ILIAS, EXO, XENON100, SuperCDMS and {\sc Majorana} collaborations.

\appendix

\section{MADF \texttt{value} arrays}

\texttt{value} arrays appear in the \texttt{measurement.results} substructure and in the \texttt{user} arrays. In both of these, the array may contain one to three numbers, whose meanings can be inferred from the measurement \texttt{type}, as described below. In addition, in the \texttt{user} arrays, the \texttt{value} array may also contain a single string.\newline

\begin{tabular}{ll}
\hline
\texttt{type, value} & Description \\
\hline

\multicolumn{2}{l}{  \texttt{measurement} } \\
\enskip\texttt{[0] }        & Central value, no error. \\
\enskip\texttt{[0,0] }     & Measurement, symmetric error. \\
\enskip\texttt{[0,0,0]}   & Measurement, asymmetric error {(+, -)}. \\
  
\multicolumn{2}{l}{  \texttt{range} } \\
\enskip\texttt{[0,0] }     & Range {(lower, upper)}. \\
\enskip\texttt{[0,0,0]}  & Range {(lower, upper)}, confidence level.  \\

\multicolumn{2}{l}{    \texttt{limit}   } \\
\enskip\texttt{[0] }     & Upper limit. \\
\enskip\texttt{[0,0]}  & Upper limit with confidence level.  \\

\multicolumn{2}{l}{    \texttt{string} } \\
\enskip\texttt{[""] }     & Meaning undefined. \\

\hline
\end{tabular}

\enskip\newline

Below, by way of illustration, is an example \texttt{results} array encoding a positive measurement of U-238 and a limit on Th-232:

\begin{verbatim}
  "results": [
    {
      "isotope": "U-238",
      "type": "measurement",
      "value": [400, 20],
      "unit": "ppb"
    },
    {
      "isotope": "Th-232",
      "type": "limit",
      "value": [100, 90],
      "unit": "ppt"
    }
  ]
\end{verbatim}

\section{MADF units}

\texttt{unit} strings appear in the \texttt{measurement.results} sub-structure and in the \texttt{user} arrays. In the \texttt{user} arrays there is no restriction on their contents but in the \texttt{measurement.results} sub-structure they must contain a value from this list: 

\begin{itemize}
\item \texttt{pct} (percent by mass)
\item \texttt{g/g}, \texttt{ppm}, \texttt{ppb}, \texttt{ppt}, \texttt{ppq} (parts per quadrillion)
\item \texttt{g}, \texttt{mg}, \texttt{ug} (micrograms), \texttt{ng}, \texttt{pg}
\item \texttt{Bq}, \texttt{mBq}, \texttt{uBq}, \texttt{nBq}, \texttt{pBq}
\item \texttt{X/Y} where \texttt{X} is any of \texttt{g}, \texttt{mg}, \texttt{ug}, \texttt{ng}, \texttt{pg}, \texttt{Bq}, \texttt{mBq}, \texttt{uBq}, \texttt{nBq}, \texttt{pBq} and \texttt{Y} is any of \texttt{unit}, \texttt{kg}, \texttt{cm}, \texttt{m}, \texttt{cm2}, \texttt{m2}, \texttt{cm3}, \texttt{m3} (where 2 and 3 represent powers)
\end{itemize}

\section{MADF dates}

\texttt{date} arrays appear in the \texttt{measurement} and \texttt{data\_source.input} sub-structures. The array can be empty or else contain one or two strings in the format \texttt{YYYY-MM-DD}, \texttt{YYYY-MM} or \texttt{YYYY}. A single string represents a date and two string represent a range of dates. Examples are:

\begin{verbatim}
  date = []
  date = ["1980-04"]
  date = ["1980-04-26", "1980-04-28"]
\end{verbatim}

\section*{References}

\bibliography{article}

\hyphenation{Post-Script Sprin-ger}
\begin{thebibliography}{10}
\expandafter\ifx\csname url\endcsname\relax
  \def\url#1{\texttt{#1}}\fi
\expandafter\ifx\csname urlprefix\endcsname\relax\def\urlprefix{URL }\fi
\expandafter\ifx\csname href\endcsname\relax
  \def\href#1#2{#2} \def\path#1{#1}\fi

\bibitem{sno}
{Q. R. Ahmad et al.},
  \href{http://link.aps.org/doi/10.1103/PhysRevLett.89.011301}{Direct evidence
  for neutrino flavor transformation from neutral-current interactions in the
  {S}udbury {N}eutrino {O}bservatory}, Phys. Rev. Lett. 89 (2002) 011301.
\newblock \href {http://dx.doi.org/10.1103/PhysRevLett.89.011301}
  {\path{doi:10.1103/PhysRevLett.89.011301}}.
\newline\urlprefix\url{http://link.aps.org/doi/10.1103/PhysRevLett.89.011301}

\bibitem{kamland}
{T. Araki et al.},
  \href{http://link.aps.org/doi/10.1103/PhysRevLett.94.081801}{Measurement of
  neutrino oscillation with {KamLAND}: Evidence of spectral distortion}, Phys.
  Rev. Lett. 94 (2005) 081801.
\newblock \href {http://dx.doi.org/10.1103/PhysRevLett.94.081801}
  {\path{doi:10.1103/PhysRevLett.94.081801}}.
\newline\urlprefix\url{http://link.aps.org/doi/10.1103/PhysRevLett.94.081801}

\bibitem{gerda}
{M. Agostini et al.},
  \href{http://link.aps.org/doi/10.1103/PhysRevLett.111.122503}{Results on
  neutrinoless double-$\ensuremath{\beta}$ decay of $^{76}\mathrm{Ge}$ from
  phase i of the {GERDA} experiment}, Phys. Rev. Lett. 111 (2013) 122503.
\newblock \href {http://dx.doi.org/10.1103/PhysRevLett.111.122503}
  {\path{doi:10.1103/PhysRevLett.111.122503}}.
\newline\urlprefix\url{http://link.aps.org/doi/10.1103/PhysRevLett.111.122503}

\bibitem{cdms}
\relax {R.~Agnese} {et al.} (SuperCDMS~Collaboration), {WIMP}-search results
  from the second {CDMSlite} run, Submitted to Phys. Rev. Lett.\href
  {http://arxiv.org/abs/1509.02448} {\path{arXiv:1509.02448}}.

\bibitem{lux}
D.~Akerib, et~al., The {L}arge {U}nderground {X}enon ({LUX}) experiment, Nucl.
  Instrum. Meth. A704 (2013) 111--126.

\bibitem{mjd2}
\relax {N.~Abgrall} {\it et al.}~({M}{\sc ajorana} Collaboration), The {M}{\sc
  ajorana} {D}{\sc emonstrator} neutrinoless double-beta decay experiment, AHEP
  (2014) 365432.

\bibitem{cuore}
{D. R. Artusa et al.}, Searching for neutrinoless double-beta decay of
  $^{130}\mathrm{Te}$ with {CUORE}, Advances in High Energy Physics\href
  {http://dx.doi.org/doi:10.1155/2015/879871}
  {\path{doi:doi:10.1155/2015/879871}}.

\bibitem{ILIAS}
{ILIAS} database on radiopurity of materials, \url{http://radiopurity.in2p3.fr}
  (accessed 2016-04-13).

\bibitem{persephone}
{P}ersephone, \url{http://github.com/radiopurity/persephone} (accessed
  2016-04-13).

\bibitem{json}
C.~Severance, Discovering {J}ava{S}cript {O}bject {N}otation, Computer 45~(4)
  (2012) 6--8.
\newblock \href {http://dx.doi.org/10.1109/MC.2012.132}
  {\path{doi:10.1109/MC.2012.132}}.

\bibitem{mapreduce}
J.~Dean, S.~Ghemawat, Map{R}educe: Simplified data processing on large
  clusters., Proc. of the 6th OSDI (2004) 137--150.

\bibitem{jsonschema}
{JSON} schema, \url{http://json-schema.org} (accessed 2016-04-13).

\bibitem{jsonvalidate}
jsonschema, \url{http://pypi.python.org/pypi/jsonschema} (accessed 2016-04-13).

\bibitem{couchdb}
Apache {C}ouch{DB}, \url{http://couchdb.apache.org} (accessed 2016-04-13).

\bibitem{lucene}
Apache {L}ucene, \url{http://lucene.apache.org} (accessed 2016-04-13).

\bibitem{couchdblucene}
couchdb-lucene, \url{http://github.com/rnewson/couchdb-lucene} (accessed
  2016-04-13).

\bibitem{cox}
\relax {G.A. Cox}~{et al.}, A multi-tiered data structure and process
  management system based on {ROOT} and {CouchDB}, Nucl. Instrum. Meth. A 684
  (2012) 63 -- 72.
\newblock \href
  {http://dx.doi.org/http://dx.doi.org/10.1016/j.nima.2012.04.049}
  {\path{doi:http://dx.doi.org/10.1016/j.nima.2012.04.049}}.

\bibitem{mjparts}
\relax {N.~Abgrall} {et al.}~({M}{\sc ajorana} Collaboration), The {M}{\sc
  ajorana} parts tracking database, Nucl. Instrum. Meth. A 779 (2015) 52 -- 62.
\newblock \href
  {http://dx.doi.org/http://dx.doi.org/10.1016/j.nima.2015.01.001}
  {\path{doi:http://dx.doi.org/10.1016/j.nima.2015.01.001}}.

\bibitem{couchapp}
couchapp, \url{http://github.com/couchapp/couchapp} (accessed 2016-04-13).

\bibitem{jquery}
j{Q}uery, \url{https://www.jquery.com} (accessed 2016-04-13).

\bibitem{jQueryUI}
j{Q}uery{UI}, \url{https://www.jqueryui.com} (accessed 2016-04-13).

\bibitem{Leonard2008490}
{D.S. Leonard et al.}, Systematic study of trace radioactive impurities in
  candidate construction materials for {EXO}-200, Nucl. Instrum. Meth. A
  591~(3) (2008) 490 -- 509.
\newblock \href {http://dx.doi.org/10.1016/j.nima.2008.03.001}
  {\path{doi:10.1016/j.nima.2008.03.001}}.

\bibitem{Aprile2012573}
{E. Aprile et al.}, The {XENON100} dark matter experiment, Astropart. Phys.
  35~(9) (2012) 573 -- 590.
\newblock \href {http://dx.doi.org/10.1016/j.astropartphys.2012.01.003}
  {\path{doi:10.1016/j.astropartphys.2012.01.003}}.

\bibitem{Alimonti1998141}
{G. Alimonti et al.}, Ultra-low background measurements in a large volume
  underground detector, Astropart. Phys. 8~(3) (1998) 141 -- 157.
\newblock \href {http://dx.doi.org/10.1016/S0927-6505(97)00050-9}
  {\path{doi:10.1016/S0927-6505(97)00050-9}}.

\bibitem{mjassay}
\relax {N.~Abgrall} {et al.}~({M}{\sc ajorana} Collaboration), The {M}{\sc
  ajorana} {D}{\sc emonstrator} radioassay program, Submitted to Nucl. Instrum.
  Meth. A\href {http://arxiv.org/abs/1601.03779} {\path{arXiv:1601.03779}}.

\end{thebibliography}

\end{document}